\documentclass[a4paper,fleqn]{cas-dc}

\usepackage[numbers,sort&compress]{natbib}

\def\tsc#1{\csdef{#1}{\textsc{\lowercase{#1}}\xspace}}
\tsc{WGM}
\tsc{QE}
\tsc{EP}
\tsc{PMS}
\tsc{BEC}
\tsc{DE}
\usepackage{epstopdf}
\usepackage{microtype}
\usepackage[T1]{fontenc}
\usepackage{amssymb}
\usepackage{hyperref}
\usepackage{changes}

\begin{document}
\let\WriteBookmarks\relax
\def\floatpagepagefraction{1}
\def\textpagefraction{.001}
\shorttitle{Tuning the rheology and microstructure of particle-laden fluid interfaces with Janus particles}
\shortauthors{Y. Qiao et~al.}

\title [mode = title]{Tuning the rheology and microstructure of particle-laden fluid interfaces with Janus particles}

\author[1]{Yiming {Qiao}}
\author[1]{Xiaolei {Ma}}
\author[1]{Zhengyang {Liu}}
\author[1]{Michael A. {Manno}}

\author[2]{Nathan C. {Keim}}
\cormark[1]
\ead{keim@psu.edu}

\author[1]{Xiang {Cheng}}
\cormark[1]
\ead{xcheng@umn.edu}


\address[1]{Department of Chemical Engineering and Materials Science, University of Minnesota, Minneapolis, MN 55455, USA}
\address[2]{Department of Physics, Pennsylvania State University, University Park, PA 16802, USA}

\cortext[cor1]{Corresponding author}

\begin{abstract}
\textit{}\textit{Hypothesis}: Particle-laden fluid interfaces are the central component of many natural and engineering systems. Understanding the mechanical properties and improving the stability of such interfaces are of great practical importance. Janus particles, a special class of heterogeneous colloids, might be utilized as an effective surface-active agent to control the assembly and interfacial rheology of particle-laden fluid interfaces.\\
\textit{Experiments}: Using a custom-built interfacial stress rheometer, we explore the effect of Janus particle additives on the interfacial rheology and microscopic structure of particle-laden fluid interfaces.\\
\textit{Findings}: We find that the addition of a small amount of platinum-polystyrene (Pt-PS) Janus particles within a monolayer of PS colloids (1:40 number ratio) can lead to more than an order-of-magnitude increase in surface moduli with enhanced elasticity, which greatly improves the stability of the interface. This drastic change in interfacial rheology is associated with the formation of local particle clusters surrounding each Janus particle. We further explain the origin of local particle clusters by considering the interparticle interactions at the interface. Our experiments reveal the effect of local particle structures on the macroscopic rheological behaviors of particle monolayers and demonstrate a new way to tune the microstructure and mechanical properties of particle-laden fluid interfaces.
\end{abstract}



\maketitle

\section{Introduction}\label{intro}

Particle-laden fluid interfaces are ubiquitous in nature and engineering applications and have received considerable research interests due to their ability to stabilize interface-rich materials like emulsions, foams and co-continuous blends \cite{Pickering1907cxcvi, ramsden1904separation, binks2005aqueous, binks2006phase, Lian2015cocontinuous,Huang2016cocontinuous}. These materials play critical roles in the processing of many industrial products such as food, cosmetics and medicine \cite{binks2002particles, hunter2008role, dickinson2010food, he2015interfacial, lotito2017approaches, rodriguez2019capsules}. Therefore, understanding and further controlling the response of particle-laden fluid interfaces to mechanical deformation are of great practical importance. Thanks to the recent advancement of interfacial rheometry, the rheology and structure of adsorbed particle monolayers at fluid interfaces are finally subjected to quantitative measurements \cite{fuller2012complex,mendoza2014particle,anjali2018shape,maestro2017nonaffine,guzman2021particle}. Monolayers of a wide range of homogeneous colloidal and nano particles have been studied, including silica nanoparticles \cite{safouane2007effect, maestro2015interfacial, zhang2016interfacial, yu2018rheology}, Ag nanoparticles \cite{krishnaswamy2007interfacial}, carbon black particles \cite{van2013rough}, poly-(methyl methacrylate) (PMMA) particles \cite{van2017interfacial}, and polystyrene (PS) particles \cite{cicuta2003shearing, reynaert2006control, reynaert2007interfacial, park2008direct, madivala2009self, barman2014simultaneous, barman2016role, beltramo2017arresting, schroyen2017versatile, rahman2019modifying} to name a few.

These pioneering studies have shown that the rheological properties of particle monolayers are closely related to their microstructures, which sensitively depend on interparticle interactions. Several methods have been invented to alter the particle-particle interaction at fluid interfaces such as modifying the chemical composition of systems (e.g. by adding salts and/or surfactants) \cite{reynaert2006control, reynaert2007interfacial, park2008direct, barman2014simultaneous, barman2016role, schroyen2017versatile, rahman2019modifying} and utilizing particles of different shapes \cite{madivala2009self} and roughness \cite{van2013rough}. These existing methods typically create percolating particle aggregates that span the entire interface, which significantly increases surface moduli and improve the stability of fluid interfaces. Nevertheless, such a drastic change in the large-scale structure is often disruptive to many properties of interface-rich materials. Moreover, changing the chemical composition may be impractical for certain systems and applications. Modification of the shape and roughness of all the particles at the interface can also be costly and poses a challenge for scale-up industrial production. Thus, it is desirable to find a new method that can tune the interfacial rheology of particle-laden interfaces via the manipulation of local particle structures with a minimum amount of additives.

Janus particles -- heterogeneous colloids composed of two or more distinct regions with different surface properties -- have been extensively studied in recent years as promising surface-active agents that can effectively modify the properties of fluid interfaces \cite{fernandez2016surface, zhang2017janus, safaie2020janus, correia2021janus, chen2021preparation}. While they have been already employed to control the collapse mechanism of fluid interfaces \cite{lenis2015mechanical, razavi2019impact} and reduce surface tension \cite{glaser2006janus, fernandez2014comparison, fernandez2014surface, fernandez2015interfacial, fernandez2018synthesis, kadam2018nanoscale, razavi2020surface}, the potential of Janus particles in modifying the rheology of fluid interfaces has not been explored heretofore.

Here, we show that the addition of a small amount of Janus particles to a homogeneous monolayer of repulsive colloidal particles can substantially increase the surface moduli and interfacial elasticity of the particle monolayer. Different from previous methods that change the large-scale structure of particle monolayers, Janus particles modify local particle configurations and maintain the homogeneous structure of fluid interfaces at large scales. Such a unique feature arises from the distinct interparticle interaction between Janus particles and homogeneous colloids at fluid interfaces. Taken together, our study provides not only a quantitative study of the interfacial shear rheology of Janus particles but also an economically viable method to tune the material microstucture and surface rheology of particle-laden fluid interfaces.

\begin{figure*}
\begin{center}
\includegraphics[width=6.8 in]{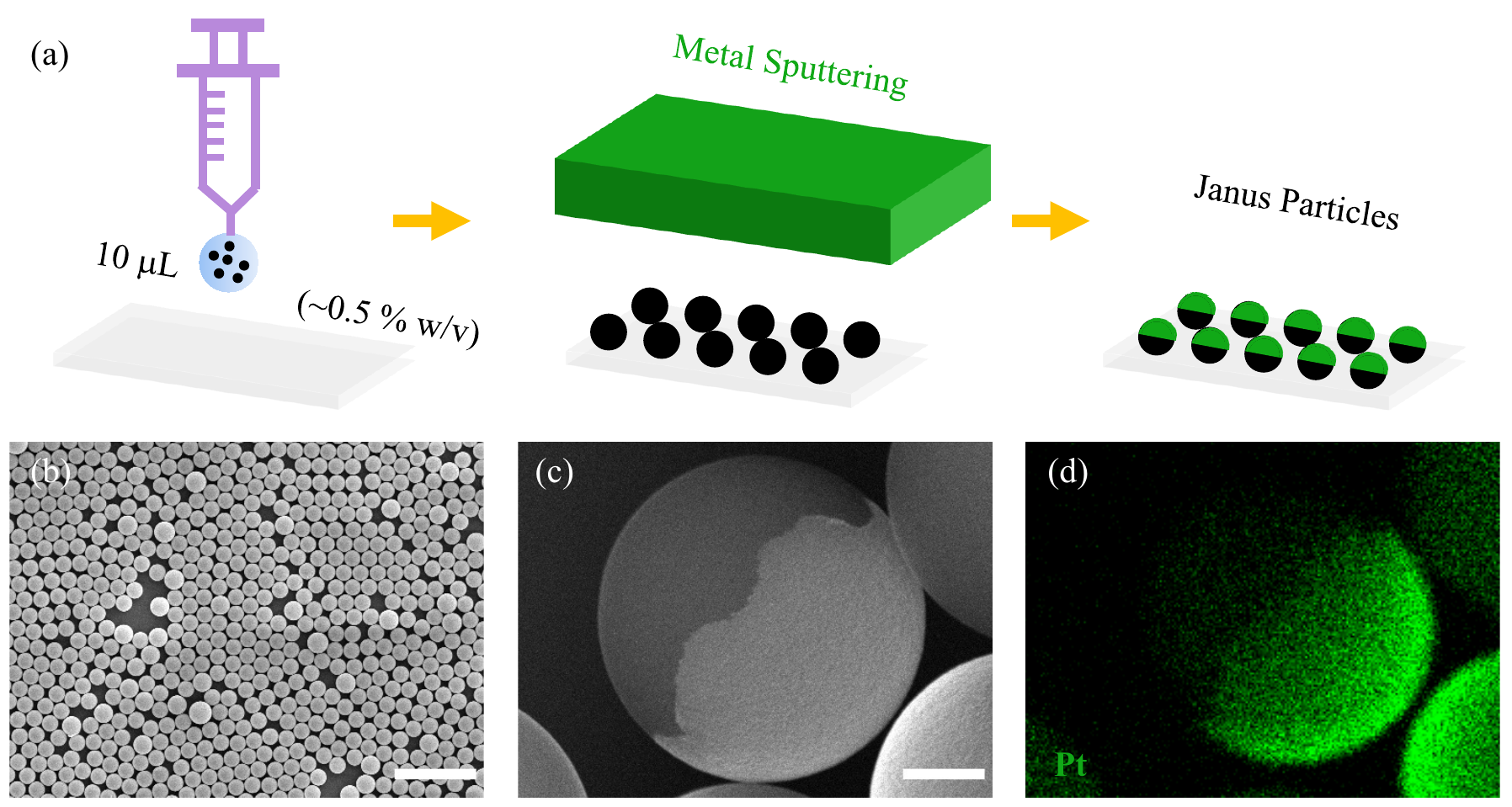}
\caption[]{Fabrication and characterization of Pt-PS Janus particles. (a) A schematic showing the protocol of Janus particle fabrication with drop casting and sputtering deposition. (b) Scanning electron microscopy (SEM) secondary electron image of a PS particle monolayer formed after drop-casting. The scale bar is 20 $\mu$m. (c) SEM image of Pt-PS Janus particles after sputtering deposition. The thickness of the coated Pt layer is $\sim$ 20 nm, estimated from deposition rate and coating time. The scale bar is 1 $\mu$m. (d) Energy dispersive x-ray spectroscopy (EDS) map of the Pt-Ps Janus particles from (c). Regions in green represent the presence of Pt.
\label{Figure_1}
}
\end{center}
\end{figure*}
%


\section{\label{Methods}Materials and Methods}

\subsection{Interfacial stress rheometer (ISR)}

Our experiments use a custom-built interfacial stress rheometer (ISR) for interfacial rheology measurements \cite{shahin1986stress,brooks1999interfacial,reynaert2008analysis,qiao2021miniature}. Briefly, a magnetic needle (79.2 $\mu$m radius, 2.45 cm length) is placed at the fluid-fluid interface between a pair of vertical glass walls. The length of the walls is 2 cm and the distance between the two walls is 6 mm. We use a pair of small permanent magnets as the magnetic trap to align the needle parallel to the glass walls. A single magnetic coil is then used to generate the perturbative field, which drives the oscillation of the magnetic needle at the fluid interface. An inverted bright-field microscope (Nikon, Ti-E) and a high-resolution CMOS camera (Basler acA2040-90um USB 3.0) are employed to image the needle motion and the structure of particles at the fluid interface. The schematic of our ISR setup is shown in Fig. S1 in the Supplementary Materials. The detailed functionality of the rheometer, the calibration process and the calculation method can be further found in our previous publication \cite{qiao2021miniature}. The validation of our measurements can be found in Fig. S2 and discussion in Supplementary Materials.

\begin{figure*}
\begin{center}
\includegraphics[width=6.8 in]{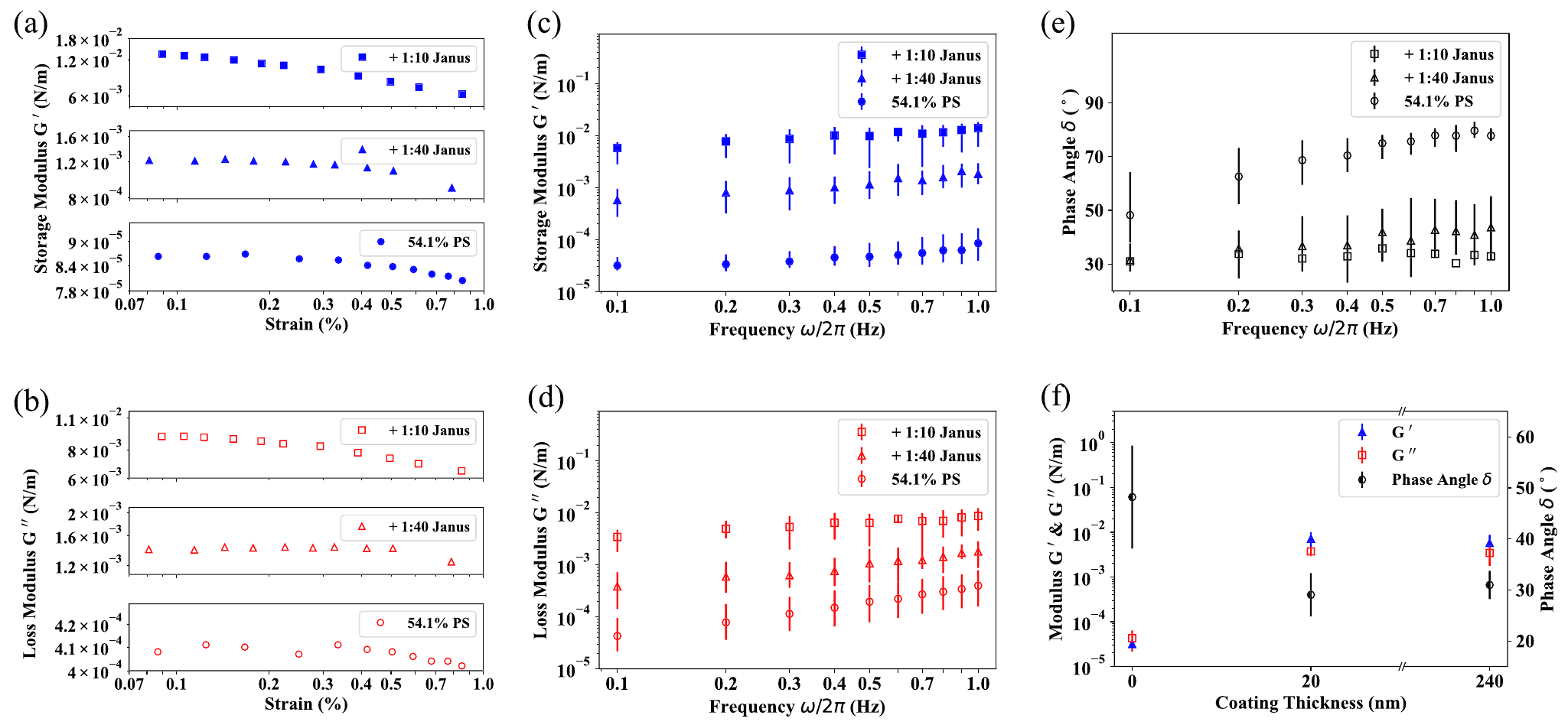}
\caption[]{Interfacial rheology of particle monolayers at the water-decane interface. The monolayers consist of PS particles with the surface area fraction of 54.1\% and different amounts of additional Pt-PS Janus particles. The number ratios between Janus and PS particles are given in the legends. (a) and (b) Storage moduli $G'$ and loss moduli $G''$ from a strain sweep, conducted at a shearing frequency of 0.5 Hz. (c) and (d) $G'$ and $G''$ from a frequency sweep, conducted at shear strains less than 0.1\% within the linear viscoelastic regime. (e) Phase angles $\delta$ from the frequency sweep. For clarity, we do not show the error bars of the 1:10 ratio sample. (f) $G'$ (closed symbols), $G''$ (open symbols) and phase angle $\delta$ (half-filled symbols) as a function of the thickness of Pt coating. Homogeneous PS particles have a thickness of 0. The number ratio between Janus and PS particles is fixed at 1:10. The shearing frequency is 0.1 Hz. The error bars in (c)–(f) indicate the ranges of quantities of multiple independent measurements.
\label{Figure_2}
}
\end{center}
\end{figure*}

\subsection{Fabrication of Janus particles}

We follow the published protocol for fabricating Janus particles using the methods of drop-casting and sputtering deposition \cite{love2002fabrication, park2011janus, dietrich2018active}. The scheme of this process is shown in Fig.~\ref{Figure_1}(a). First, 10 $\mu$L dilute aqueous suspension ($\sim 0.5\%$ w/v) of 4.1-$\mu$m-diameter polystyrene (PS) spheres is spread onto a glass slide and dried at room temperature. The PS particles were purchased from Thermo Fisher Scientific (lot No. 1876103), which are sulfate-modified and surfactant-free. Due to the low concentration of the spreading suspension, the particles form a single layer on the glass substrate, as characterized by scanning electron microscopy (SEM) in Fig.~\ref{Figure_1}(b). We then deposit thin films of platinum (Pt) on the upper surface of the PS particles with sputtering deposition (AJA International, Inc. ATC 2000), a method based on physical vapor deposition. At a given deposition rate, the coating thickness is estimated from the coating time based on a calibration curve.

Afterwards, the Pt-coated PS particles are released from the glass substrate into deionized (DI) water (Millipore Direct-Q3, 18.2 $\mathrm{M} \Omega \cdot \mathrm{cm}$ at 25${ }^{\circ} \mathrm{C}$) by sonication for approximately 1 min. The resulting suspension is centrifuged to remove the supernatant. The particles are then washed with DI water and reagent grade alcohol (100\%, VWR International) six times. Figure ~\ref{Figure_1}(c) shows the SEM image of the fabricated Pt-PS Janus particles with a Pt coating thickness of 20 nm. We further verify the presence of Pt with energy dispersive X-ray spectroscopy (EDS), as shown in Fig.~\ref{Figure_1}(d). Pt-PS Janus particles of different coating thicknesses show similar structures.

    \subsection{Preparation of particle-laden fluid interfaces}
We first prepare seperate suspensions of PS particles and Pt-PS Janus particles. To make the suspensions with the same particle number concentration, particle suspensions are first dried in an oven at 60 $^\circ$C. The number of particles is calculated by using the size and the density of particles and the total mass of particles, which is obtained from the weight gain of the container. The density of the PS particles is 1.055 g/${\mathrm{cm}^3}$, given by the manufacturer. We estimate the density of the Janus particles by assuming a hemisphere of Pt coating on each particle. The particle number concentration of the suspensions can be adjusted by adding a certain amount of the mixture of reagent grade alcohol (50\% w/v) and distilled DI water (50\% w/v). Finally, we mix the suspension of PS particles with different amounts of the suspension of Janus particles to achieve the desired number ratio of Janus/PS particles (1:40 or 1:10) in the final spreading suspension.

We prepare a clean water-decane interface by adding 10 mL of water as the subphase and 10 mL of decane (Sigma-Aldrich) as the superphase. The magnetic needle of ISR is gently placed at the interface with tweezers. After that, we slowly deposit the well-sonicated spreading suspension at the clean water-decane interface to form an interfacial particle monolayer, which remains stable due to the high detachment energy \cite{pieranski1980two,binks2006phase,hunter2008role,guzman2021particle}. By controlling the volume of the spreading suspension, we vary the area fraction of particles at the interface from $\sim$ 30.0\% up to $\sim$ 50.0\% with an error about $\sim$ 4.6\% between different runs (Fig. S3). After the particles reach equilibrium at the interface ($\sim$ 10 min), we apply large shear strains on the interface for 10 min. When no particle motions are further observed ($\sim$ 10 min after stopping the shear), the sample is ready for interfacial rheological measurements. We visually inspect the particle structure at different locations of the interface and verify that the interface has a uniform structure with the desired area fraction and Janus-PS particle ratio before each experiment.

\section{\label{Results}Results and Discussion}

\begin{figure*}
\begin{center}
\includegraphics[width=6 in]{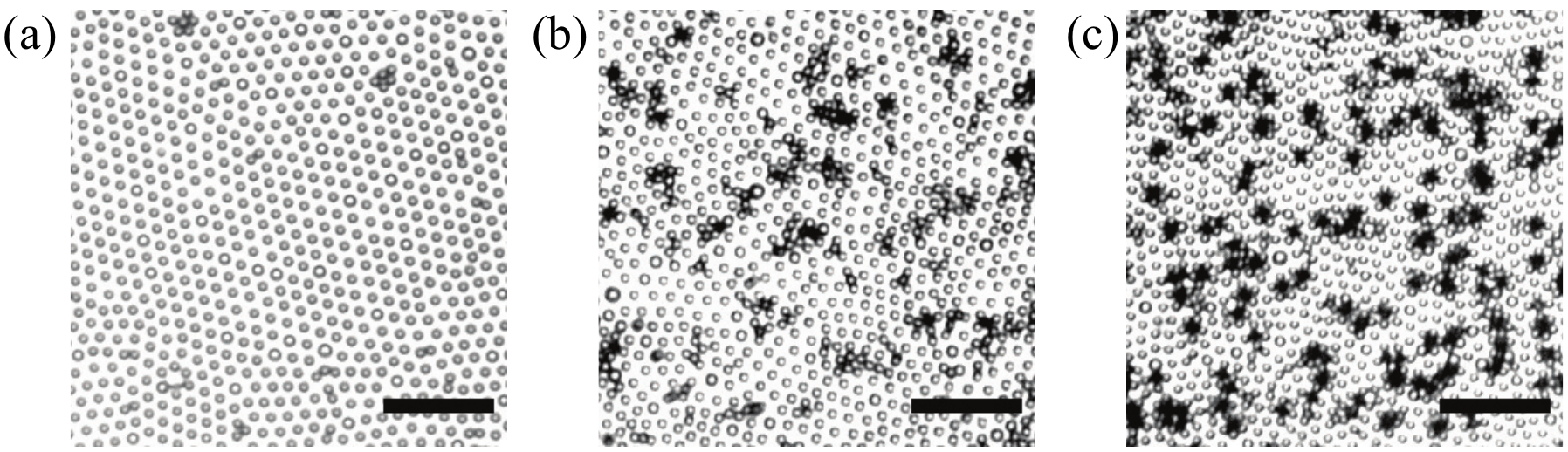}
\caption[]{Bright-field microscopy images of particle monolayers at the water-decane interface. (a) A monolayer of homogeneous PS particles. (b) A monolayer with 1:40 number ratio of Janus and PS particles. (c) A monolayer with 1:10 number ratio of Janus and PS particles. The area fractions of PS particles are kept around 38\% in all three monolayers. The dark particles in (b) and (c) are Janus particles, whose Pt coating layers have a thickness of 20 nm. The scale bars are 50 $\mu$m.
\label{Figure_3}
}
\end{center}
\end{figure*}

\subsection{Interfacial rheology}

We first examine the linear viscoelasticity of particle monolayers of different compositions. Figures \ref{Figure_2}(a) and \ref{Figure_2}(b) show the storage and loss modulus, $G'$ and $G''$, of particle monolayers with the area fraction of PS particles at 54.1\% and different amounts of Janus particles, in a strain sweep at a fixed shearing frequency of 0.5 Hz. The strain amplitude, i.e., the amplitude of the oscillatory displacement of the needle divided by the sample dimension between the needle and the wall, ranges between 0.08\% and 1\% in our experiments. The linear viscoelastic region exists when the strain amplitude is less than 0.1\% for all the three samples tested, which is consistent with previous studies of the interfacial rheology of homogeneous particle monolayers \cite{reynaert2007interfacial, barman2016role, qiao2021miniature}.

Within the linear viscoelastic regime, Figures \ref{Figure_2}(c) and \ref{Figure_2}(d) further show $G'$ and $G''$ in a frequency sweep with shearing frequencies ranging between 0.1 and 1.0 Hz. We observe a drastic increase of $G'$ and $G''$ when only a small amount of Pt-PS Janus particles are added to the PS particle monolayers. Specifically, $G'$ and $G''$ show an increase of more than one order of magnitude when the number ratio of Janus particles to PS particles is 1:40, and show an increase of more than two orders of magnitude when the ratio increases to 1:10. Furthermore, the phase angles, $\delta \equiv \arctan(G''/G')$, of the Janus/PS particle monolayers are consistently smaller than those of the homogeneous PS particle monolayer across all frequencies tested (Fig.~\ref{Figure_2}(e)). Hence, the presence of Pt-PS Janus particles renders particle monolayers more elastic. Particularly, at the 1:10 Janus/PS particle number ratio, $\delta$ decreases more than $40^\circ$ at high shear frequencies, indicating a substantial increase of interfacial elasticity. Both the strong increase in $G''$ and the greatly enhanced elasticity improve the stability of particle monolayers at the fluid interface. Note that the above findings do not depend on the thickness of the Pt coating of Janus particles. Nearly the same results are obtained when we change the coating thickness from 20 nm to 240 nm (Fig. \ref{Figure_2}(f)).

It should be emphasized that the drastic increases of $G'$, $G''$ and interfacial elasticity cannot be simply attributed to the increase of the area fraction of the particle monolayers due to the addition of Janus particles. For an increase of one order of magnitude in $G'$ and $G''$, the area fraction of homogeneous PS particles needs to increase by $\sim$ 10\% (Fig. S4(a) and (b)), which is far larger than the 1\% increase in area fraction at the number ratio of Janus/PS particles of 1:40. Similarly, an increase of two orders of magnitude in $G'$ and $G''$ is observed when the area fraction of homogeneous PS particle monolayers increases by $\sim$ 17\% (Fig. S4(a) and (b)), much larger than the 5\% increase in area fraction at the Janus/PS particle number ratio of 1:10. More importantly, the phase angle $\delta$ decreases only slightly, almost negligible at low shear frequencies, for homogeneous PS particle monolayers of increasing area fractions (Fig. S4(c)) \cite{qiao2021miniature}. Thus, the pronounced increase in the surface moduli and the accompanying enhancement of interfacial elasticity must be associated with the unique properties of Janus particles. The presence of Janus particles profoundly affects the interparticle interaction and modifies the microscopic structure of particle monolayers, as we shall show in the next section.

As an additional comment, the enhancement of shear modulus due to Janus particles allows monolayers with even relatively small area fractions to be successfully studied in our rheometer. For homogeneous PS particle monolayers, when the area fraction is below 45.6\%, the surface moduli are too small to measure reliably by our ISR rheometer. At such low area fractions, significant momentum transfer occurs between the interface and the bulk fluids, which results in nonlinear velocity profiles of particles invalidating the simple calculation method of our ISR (Fig. S5 and the discussion in Supplementary Materials). In contrast, with the addition of a small amount of Janus particles, the velocity profiles of particles remain linear even at a low surface area fraction of $\sim$ 40\% (Fig. S5), which justifies our calculation method and further supports the key conclusion of our experiments, i.e., Janus particles drastically enhance surface moduli and elasticity of particle monolayers.

\subsection{Microstructure}

\begin{figure*}
\begin{center}
\includegraphics[width=6.8 in]{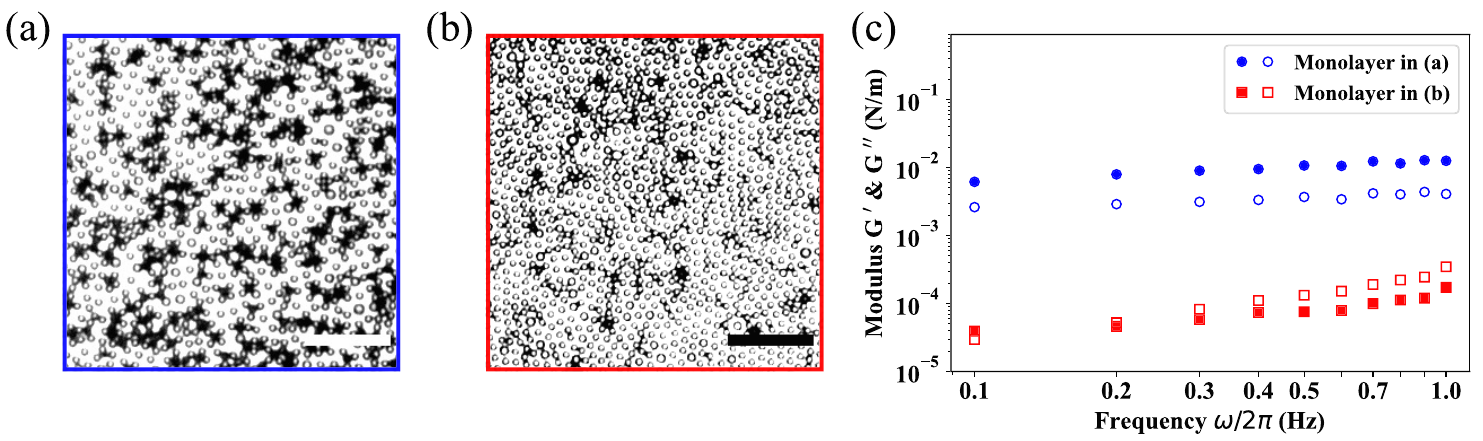}
\caption[]{A comparison of the structure and rheology of two particle monolayers of different compositions at the water-decane interface. (a) A microscopy image of a monolayer of Janus/PS particles, where the area fraction of PS particles is $\sim$ 25\%. The number ratio between Janus and PS particles is 1:5. (b) A microscopy image of another monolayer of Janus/PS particles, where the area fraction of PS particles is $\sim$ 40\%. The number ratio between Janus and PS particles is 1:50. The scale bars are 50 $\mu$m. (c) A frequency sweep in the linear viscoelastic regime of the storage and loss moduli $G'$ and $G''$ of the two particle monolayers shown in (a) and (b). Closed symbols are for $G'$ and open symbols are for $G''$.
\label{Figure_4}
}
\end{center}
\end{figure*}

To explore the origin of the enhanced interfacial rheology, we image directly the microscopic structure of the particle monolayers. Figure~\ref{Figure_3} shows the structure of a homogeneous PS particle monolayer together with the structure of particle monolayers with 1:40 and 1:10 number ratio of Janus and PS particles. To better illustrate the microstructure at the single particle level, we lower the area fraction of PS particles in the monolayers to $\sim$ 38\%. Nevertheless, qualitatively similar features have also been observed at higher area fractions, (Fig. S6). The monolayer of charge-stabilized homogeneous PS particles shows a hexagonal crystalline structure with a lattice constant about two particle diameters, consistent with previous studies \cite{reynaert2006control,reynaert2007interfacial,park2008direct,dietrich2018active,pieranski1980two,park2010heterogeneity}. In comparison, the presence of Janus particles induces local particle clusters in the particle monolayers, where each Janus particle is bonded with 3--5 PS particles. Due to the low concentration of Janus particles, the clusters are compact and small at the scale of a few particles, which are in sharp contrast to the system-sized percolating fractal particle aggregates observed in previous studies \cite{van2013rough,reynaert2006control,reynaert2007interfacial,park2008direct,madivala2009self,barman2014simultaneous,barman2016role}. The formation of local particle clusters is also independent of the thickness of the Pt coating of Janus particles within the range of our experiments from 10 nm up to 240 nm (Fig. S7), agreeing with our rheology measurements (Fig. 2(f)). Such local clusters induced by the attraction between Janus particles and unmodified PS particles are similar to the ``flower-like'' structure reported in Ref. \cite{kozina2018bilayers}, where bilayers of homogeneous hydrophobic particles and amphiphilic Janus particles were created at a water-air interface.

The formation of local clusters leads to the pronounced increase of the surface moduli and elasticity. To further demonstrate the correlation between particle clusters and the enhanced interfacial rheology, we compare two more particle monolayers (Fig.~\ref{Figure_4}). Figure~\ref{Figure_4}(a) shows a monolayer of PS particles of a low surface coverage of $\sim$ 25\% with a high number ratio of Janus and PS particles of 1:5, while the monolayer in Fig. \ref{Figure_4}(b) has a higher PS particle surface coverage of $\sim$ 40\% but a much lower number ratio of Janus and PS particles of 1:50. As the number of particle clusters is proportional to the number of Janus particles, the monolayer in Fig.~\ref{Figure_4}(a) has $\sim$ 6 times more particle clusters than the monolayer in Fig.~\ref{Figure_4}(b), although the overall area fraction of the former is about 1.4 times smaller. Strikingly, the surface moduli of the less concentrated monolayer in Fig.~\ref{Figure_4}(a) are two orders of magnitude higher than those of the more concentrated monolayer in Fig.~\ref{Figure_4}(b) (Fig.~\ref{Figure_4}(c)). This set of control experiments unambiguously demonstrates that the concentration of particle clusters dominates over the total concentration of particles in determining the rheological response of particle-laden fluid interfaces.

To explain the physical origin of Janus-PS particle clusters, we consider the interplay between the two dominant interparticle interactions at the interface, i.e., dipolar electrostatic repulsion and capillary attraction \cite{dietrich2018active,park2011attractive,bresme2007nanoparticles,oettel2008colloidal,dani2015hydrodynamics,garbin2012nanoparticles}. The dipolar repulsion is due to the asymmetric charge distribution of particles at the interface between water and a nonpolar fluid such as decane \cite{park2008direct,pieranski1980two,park2010heterogeneity,aveyard2002measurement,danov2010interaction}, which leads to a repulsive potential:
\begin{equation}
\frac{U_{\text {dipole }}}{k_{B} T}=\frac{a_{p p}}{r^{3}}.
\end{equation}
Here, $k_{B} T$ denotes the thermal energy, where $k_{B}$ is the Boltzmann constant and $T$ is the room temperature. $a_{p p}$ is a prefactor that quantifies the magnitude of the repulsive interaction and $r$ is the center-to-center distance between particles. For sulfate-modified PS particles at the water-decane interface, $a_{p p}$ is $\sim$ $10^{-13}$ m$^{3}$ \cite{park2010heterogeneity,park2011attractive,masschaele2010finite}. Metal coating modifies the charge distribution of particles, which significantly lowers the strength of repulsion. Ref.~\cite{dietrich2018active} shows that $a_{pp}$ can be three orders of magnitude smaller for Pt-PS Janus particles. Thus, we take $a_{pp} \sim10^{-16}$ m$^{3}$ for our Janus particles. The capillary attraction is ascribed to the contact line undulations induced by particle surface roughness \cite{danov2005interactions,botto2012capillary,liu2018capillary}. The corresponding potential between particles A and B of radius $R$ at the fluid-fluid interface is:
\begin{equation}
U_{\text {capillary}}=-\frac{12 \pi \gamma H_{A} H_{B} R^{4}}{r^{4}},
\end{equation}
where $\gamma=$ 50 $\mathrm{mN} / \mathrm{m}$ is the water-decane surface tension, $H_{A}$ and $H_{B}$ correspond to the surface roughness of the particles \cite{dietrich2018active}. We choose $H_{A} = 2$ nm as the surface roughness of PS particles \cite{dietrich2018active} and $H_B = 2h + H_A$ as the effective surface roughness of Pt-PS Janus particles, where $h$ is the thickness of the Pt coating layer \cite{dietrich2018active}. Note that the conclusion of our study below does not depend on the exact value of $H_A$. Any choice of $H_A$ on the order of 1 nm would lead to qualitatively the same results.

\begin{figure*}
\begin{center}
\includegraphics[width=6 in]{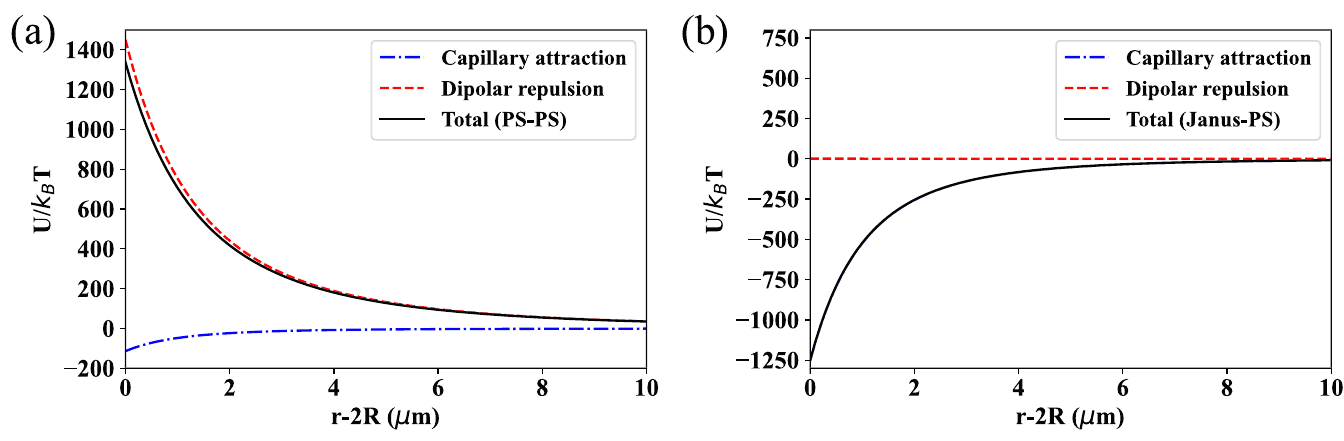}
\caption[]{Interparticle potentials between a pair of PS particles (a) and between a Janus particle and a PS particle (b). The potentials are plotted as a function of the surface-to-surface distance r{-}2R. Red dashed lines represent dipolar electrostatic repulsion. Blue dash-dotted lines show capillary attraction. Black solid lines are the total potential. Note that the capillary attraction dominates the interparticle interaction between the Janus particle and the PS particle in (b). Thus, the total potential is nearly the same as that of the capillary attraction, and the curves are indistinguishable.
\label{Figure_5}
}
\end{center}
\end{figure*}

Figure~\ref{Figure_5}(a) shows the sum of the two potentials for homogeneous PS particles, where the dipolar repulsion dominates the interparticle interaction. Such a strong repulsion gives rise to the crystalline structure observed in homogeneous PS particle monolayers (Fig.~\ref{Figure_3}(a)). With the reduction of $a_{pp}$, the dipolar repulsion becomes much weaker for Pt-PS Janus particles. As a result, even at the thinnest Pt coating of $h = 10$ nm studied in our experiments, the capillary attraction overcomes the dipolar repulsion, resulting in a large negative potential minimum of $\sim$ 1200$k_BT$ at the contact of Janus and PS particles (Fig.~\ref{Figure_5}(b)). This strong attraction between Janus and PS particles leads to the formation of local particle clusters in the monolayers of Janus/PS particles (Figs.~\ref{Figure_3}(b) and (c), Fig. S6 and Fig. S7). With further increasing $h$ and therefore effective surface roughness, the attraction becomes even stronger. Hence, the competition between the eletrostatic repulsion and the capillary attraction explains the qualitative difference between the microscopic structure of the monolayers of homogeneous PS particles and that of the monolayers of Janus/PS particles at the water-decane interface. This difference in microstructures consequently leads to the drastically different interfacial rheology of the two types of particle monolayers. It is worth of noting that although the thickness of the coating layer controls the bonding strength between Janus and PS particles in clusters, it does not affect the structure of particle clusters. Janus-PS particle clusters are always small with an average size of 3--5 particles, independent of $h$ within the range of our experiments (Fig. S7). Our experiments suggest that for the range of $h$ and shear conditions studied here, the Janus-PS bonds are so strong that each Janus-PS cluster acts as a rigid inclusion in the monolayer and is not deformed by shear. Hence the Janus-PS bond strength does not affect the interfacial rheology (Fig.~\ref{Figure_2}f).

\section{Conclusion and Outlook}

We have investigated the effect of platinum-polystyrene (Pt-PS) Janus particle additives on the rheology and structure of the monolayers of polystyrene (PS) particles at the water-decane interface. With the addition of a small amount of Janus particles, the surface moduli and elasticity of the monolayers are substantially improved. This drastic change in rheology is correlated with the unique microscopic structure of Janus/PS particle monolayers, where each Janus particle attracts a small number of surrounding PS particles and form local compact clusters. These particle clusters arise from the competition between dipolar electrostatic repulsion and capillary attraction. As such, our finding opens up a new way to tune the mechanical properties and enhance the stability of particle-laden fluid interfaces. As only a small amount of Janus particles are needed in this approach, our results overcome the difficulty associated with the scale-up production of Janus particles and facilitate the potential use of Janus particles in industrial applications.

Although our experiments provide unambiguous evidence for the correlation between the enhanced rheology and the formation of particle clusters, it is still an open question how the local compact particle clusters transform the macroscopic rheology of particle-laden fluid interfaces. Different from previous studies showing system-wide particle aggregations \cite{van2013rough,reynaert2006control,reynaert2007interfacial,park2008direct,madivala2009self,barman2014simultaneous,barman2016role}, where the percolating particle structure provides the backbone supporting the shear force across the system, the shear force cannot propagate via the non-percolating particle clusters in our systems. A possible explanation is that the imposed shear organizes particle clusters into force chains along its compressive axis. Such a fragile anisotropic hierarchical structure can then enhance the surface moduli and elasticity of the fluid interface, similar to its effect on the 3D counterparts of colloidal suspensions \cite{cates1998fragile}. Further analyses on the dynamic structure of particle monolayers under shear are needed to verify the hypothesis.

The surface chemistry of Janus particles may also affect the structure and interfacial rheology of the particle monolayers \cite{correia2021janus}. Ref. \cite{park2011janus} shows that the interactions between Janus particles can be tuned by modifying the coated surface with appropriate chemicals. Ref. \cite{razavi2019impact} further demonstrates that Janus particles with different surface amphiphilicity---the wettability difference of the two hemispheres---can result in different interfacial behaviors of particle monolayers under compression. How the surface chemistry of Janus particles influences the interfacial rheology of particle-laden fluid interfaces is an interesting topic for future studies.

Lastly, when submerged in hydrogen peroxide solutions, Pt-PS Janus particles become active and show self-propelling motions by catalyzing the decomposition of hydrogen peroxide \cite{golestanian2005propulsion, golestanian2007designing, howse2007self}. When such active particles are embedded in 3D colloidal gels, they decrease the linear viscoelastic moduli and yield stresses of the gels \cite{Szakasits2019colloidalgels,Saud2021yield}. The activity of Janus particles also likely reduces the surface moduli and elasticity of fluid interfaces, an intriguing effect that can provide a further control over the behaviors of fluid interfaces and is currently under our investigation.

\section*{Acknowledgements}
We thank Chen Fan, Yangming Kou, and Sungyon Lee for the help with experiments and fruitful discussions. The research is supported by the U.S. National Science Foundation (Nos. 1700771 and 2032354). Portions of this work were conducted in the Minnesota Nano Center, which is supported by the National Science Foundation through the National Nanotechnology Coordinated Infrastructure (NNCI) under Award Number ECCS-2025124.

\printcredits

\bio{}
\endbio

\end{document}